# Correlated enhancement of $H_{c2}$ and $J_c$ in carbon nanotube-doped $MgB_2$

A. Serquis[1], G. Serrano[1], S. Moreno[1], L. Civale[2], B. Maiorov[2], F. Balakirev[3], M. Jaime[3]

[1] Centro Atómico Bariloche-Instituto Balseiro, R8402AGP, S.C.de Bariloche, ARGENTINA
[2] Superconductivity Technology Center, MS K763, Los Alamos National Laboratory, Los Alamos, NM 87545, USA
[3] NHMFL at Los Alamos National Laboratory, MS E536, Los Alamos, NM 87545, USA

E-mail: aserquis@cab.cnea.gov.ar

**Abstract.** The use of $MgB_2$ in superconducting applications still awaits for the development of a $MgB_2$-based material where both current-carrying performance and critical magnetic field are optimized simultaneously. We achieved this by doping $MgB_2$ with double-wall carbon nanotubes (DWCNT) as a source of carbon in polycrystalline samples. The optimum nominal DWCNT content for increasing the critical current density, $J_c$ is in the range 2.5-10%at depending on field and temperature. Record values of the upper critical field, $H_{c2}(4K) = 41.9$ T (with extrapolated $H_{c2}(0) \approx 44.4$ T) are reached in a bulk sample with 10%at DWCNT content. The measured $H_{c2}$ vs T in all samples are successfully described using a theoretical model for a two-gap superconductor in the dirty limit first proposed by Gurevich et al.

The superconductor $MgB_2$ has large potential for technological applications due to its high critical temperature Tc (~39K) [1], low cost of raw materials, chemical simplicity and absence of weak-links limitations to the critical current density [2]. In the last few years, several groups have achieved a good improvement of the transport properties of this material, especially in thin films [3],[4],[5],[6]. However, in polycrystalline $MgB_2$, one of the most important remaining challenges is to increase the upper critical field ($H_{c2}$), which in the case of undoped clean material is too low for most possible applications [7] while, at the same time, improving the critical current density ($J_c$). The best results for increasing $J_c$ and irreversibility field ($H_{irr}$) in bulk samples are related to an improvement in grain connectivity [5] but also to the addition of suitable defect nanoparticles or doping, i.e. $Mg(B_{1-x}O_x)_2$ [8], SiC [6],[9], Al [10], $Dy_2O_3$ [11], and carbon nanotubes (CNT) [12],[13],[14]. It is well known that pinning of vortex lines to defects in superconductors plays an extremely important role in determining their properties. CNT inclusions, with diameters close to the $MgB_2$ coherence length ($\xi_{ab}(0)$~3.7-12nm; $\xi_c(0)$~1.6-3.6 nm) [7] may be very good candidates for vortex pinning if they do not completely dissolve in the matrix but remain as tubes acting as columnar defects. In particular, Dou *et al* [12] reported an enhancement of $J_c$ by doping with multi-wall CNT controlling the extent of C substitution during the synthesis (changing sintering time and temperature) or varying the diameter and length of CNT [13]. However, these studies were made only with samples with a nominal composition $MgB_{1.8}C_{0.2}$.

Another interesting issue is that theoretical models predict that the presence of two superconducting gaps could allow tuning of different upper critical fields by controlling diverse defect sublattices relative to orthogonal hybrid bands [15],[16]. These models expect a significant $H_{c2}$ enhancement in the dirty limit and an anomalous $H_{c2}(T)$ upward curvature. Several reports indicate that $H_{c2}$ can be significantly increased by introducing disorder through oxygen alloying, carbon doping or He-ion irradiation [4],[17],[18],[19]. All the record $H_{c2}$ values are reported for films or fiber-textured samples (i.e. Braccini *et al* [4] reported $H_{c2}^{\perp}(4.2) \sim 35$ T and $H_{c2}^{\parallel}(4.2) \sim 51$ T, observed perpendicular and parallel to the *ab* plane, respectively, in epitaxial $MgB_2$ C-alloyed films and fiber-textured samples). In carbon-doped *bulk* $MgB_2$ samples much lower extrapolated $H_{c2}(0)$ values of between 29 and 38 T have been reported [20],[21],[22],[23].

Although the effect of carbon substitution was one of the most studied in $MgB_2$, the results on C solubility and the effect of C-doping on the critical temperature ($T_c$) and critical current density ($J_c$) reported, so far, vary significantly due to precursor materials, fabrication techniques and processing conditions used, leading to different levels of C substitution and possibly other impurity effects [[20],[21],[22],[23],[24]].

In this work CNT doped $MgB_2$ samples were prepared by solid-state reaction using as starting materials amorphous boron powder (-325 mesh, 99.99% Alfa Aesar), magnesium powder (-325 mesh, 99.8% Assay) and double-walled carbon nanotubes DWCNT (diameter 1.3-5 nm, length ≤ 50 μm, 90% Aldrich). Details of the reaction procedure will be reported elsewhere [[25]]. TEM observations were made to characterize the initial composition, main impurities, average CNT sizes, and morphology of the Aldrich DWCNT powder that may affect the pinning properties. The typical DWCNT diameter size of ~3 nm (see lower inset of Fig. 1) may generate defects to act as effective flux pinning centers. Of the CNTs observed, ~90% of these were DWCNT, with the remaining additives being a variety of other kinds of CNTs and graphite onion structures, with diameters as large as 15 nm. Amorphous C is also present in the sample and this probably tends to dissolve easily within $MgB_2$ structure. The shift in the a-lattice parameter, obtained from measured x-ray diffraction patterns, can be used as a measure of the actual amount of C (x) in the $Mg(B_{1-x}C_x)_2$ structure [[21]]. The x values obtained from using the fitting of Avdeev *et al.*'s neutron diffraction data [[24]] and Kazakov *et al* single crystal data [[26]] are listed in Table I. It seems that all C is incorporated into the $MgB_2$ structure for samples with nominal content lower than 5%at, while some C does not dissolve for larger nominal contents indicating that some CNT remains as nanotubes. An increase of the broadening of the peaks is observed with increasing C in the samples, which may be related to a larger internal strain [[27]].

$T_c$ values as determined by magnetization ($T_c^{mag}$) and resistivity data ($T_c^{trans}$) are plotted as a function of the actual C content (x) in Fig. 1. The $\Delta T_c$ (90-10%) of the superconducting transitions are lower than 1K in all cases indicating a homogeneous distribution of the C incorporated into the lattice. The $T_c(x)$ dependence is similar to other reported data [[21],[26]], but $T_c$ values are slightly lower in our samples due to an increase in the lattice strain [[27]]. We observe an increase in the normal state resistivity with x (from 9 to 200μΩ.cm) that is indicative of the shift to a dirty limit. As a consequence, a continuous decrease in the RRR values (defined as the ratio of resistivities $\rho(300K)/\rho(40K)$) is observed, as shown in the upper inset of Fig. 1. If we take into account that values of $\rho \sim 0.4\text{-}1.6$ µΩ.cm were used as typical of $MgB_2$ films in the clean limit [[4]], even the x = 0 sample (9 µΩ.cm) is within the dirty limit. However, there is no simple correlation between the normal state $\rho$ and $H_{c2}$, because the global resistivity may be limited by poor intergrain connectivity while $H_{c2}$ is controlled by intragrain impurity scattering.

The $J_c$'s and their field dependence calculated from the Bean model [[28]] are shown in Fig. 2 for 5 and 20K. For clarity only a few samples are included. CNT increases the amount of pinning centers, being the optimum doping temperature and field dependent. This increase may be coming in part from pinning of the remaining CNT's and part from C doping. The inset in Fig. 2 illustrates the $J_c$ performance at 4T of several samples as a function of x. For nominal contents between 2.5 – 10%at we find that $J_c$ is increased up to $5 \times 10^4$ A/cm$^2$ at 5T and 5K. It is apparent that $J_c$ decreases for x > 0.05 (nominal content larger than 10%at) due to both a larger decrease in $T_c$ and a probable deterioration of interconnectivity between grains denoted by a large $\rho(40)$ value (~ 200 µΩ.cm) for the CNT125 sample.

The $H_{c2}(T)$ dependence were determined from four-probe transport measurements in the mid-pulse magnet of NHMFL-LANL, capable to generate an asymmetric field pulse up to 50T, performed at temperatures between 1.4 and 34 K. Fig. 3(a) exhibits the temperature dependence of $H_{c2}$ and $H_{irr}$, defined as the onset (extrapolation of maximum slope up to normal state resistivity) and the beginning of the dissipation, respectively, of the R vs H data of samples with several CNT contents. It is apparent in these $H_{c2}(T)$ data the upward curvature signaled as a characteristic of the presence of two gaps [[16],[19]]. The inset displays $H_{c2}$ as function x for 4 and 20K and the extrapolation to 0K for all samples. We observe that $H_{c2}(0)$ has a maximum near x ≈ 0.045 for sample CNT10 and a decrease above 0.045 similar to the $J_c$ behavior. This enhancement is indicating that the incorporation of C in the lattice affects the scattering mechanism, consistent with the RRR variation with x, increasing $H_{c2}$ up to 44 T at 0K, as predicted by the theoretical models. A lower $H_{c2}$ enhancement is observed in single-wall CNT doped samples [[25]].

We used eq. 1 from ref. [[16]], obtained from the Usadel equations for a two-gap superconductor in the dirty limit (from Gurevich *et al* [[15],[19]] and Golubov *et al* [[16]]) to describe the $H_{c2}$ dependence on temperature observed in CNT doped samples. The proposed model considers that the nonmagnetic impurities affect the intraband electron diffusivities $D_\sigma$ and $D_\pi$, and the interband scattering rates $\Gamma_{\pi\sigma}$ and $\Gamma_{\sigma\pi}$. We optimized the diffusivity ratio $\eta = D_\pi / D_\sigma$ and interband scattering parameter $g = (\Gamma_{\sigma\pi} + \Gamma_{\pi\sigma}) / 2\pi k_B T_{c0}$, where $T_{c0} = T_c(g=0)$ to fit the measurements using the following equation [4]

$$2w(\ln t + U_+)(\ln t + U_-) + (\lambda_0 + \lambda_i)(\ln t + U_+) + (\lambda_0 - \lambda_i)(\ln t + U_-) = 0 \qquad \textbf{(eq.1)}$$

where $t = T/T_{c0}$ is the reduced temperature, $U_\pm = U_\pm(T, H_{c2}, D_\sigma, D_\pi, \Gamma_{\pi\sigma}, \Gamma_{\sigma\pi})$, $\lambda_i = \lambda_i(\Gamma_{\pi\sigma}, \Gamma_{\sigma\pi})$ and $w$, $\lambda_0$ are constants that depend on $\lambda_{mn}$ ($m,n = \pi,\sigma$) values obtained from *ab initio* calculations [4],[15],[16].

Fig. 3(b) shows $H_{c2}(T)$ as a function of the reduced temperature $t$, including the experimental values and the curves obtained by fitting the experimental data with the theoretical model of eq.(1), for CNT00, CNT01 and CNT10 samples. The upward curvature, characteristic of a two gap effect, is clearly observed near $t \approx 0.2$ for the CNT10 sample. The data for all samples are fitted with the same equation, where a clear difference between them can be explained as an effect of a change in the scattering mechanism into the bands. This indicates that the two gap nature is preserved after carbon-doping, consistent with previous measurements [4],[20]. The fitting parameters $\eta$ and $g$ are listed in Table I for all samples. The $g$ values increase with x, as the $T_c$ of the samples decreases. It is worth to note that, although $\eta$ does not follow a clear tendency, we observe that $1/D_\pi$ follows the same dependence of $H_{c2}(0)$ as a function of x (see inset of Fig. 3(b)) signifying that C-doping is affecting the π-band.

In summary, we prepared samples doped with DWCNT and determined that the actual amount of C in the lattice is less than the nominal CNT content. The decrease in $T_c$ may be explained by assuming not only C incorporation to the lattice but also an increase in the lattice strain. DWCNT doping produced the two desired effects: improvement of $J_c$ and $H_{c2}$ for 0.02 < x < 0.05. The optimum CNT content for increasing $J_c$ depends on H and T, and it may be further improved by controlling the amount of CNT in the grain boundaries.

$H_{c2}(T)$ can be described by the two-bands model[15],[4] for a dirty two gaps superconductor that takes into account the interband scattering. $H_{c2}$ in $MgB_2$ bulk samples may be increased up to record value $H_{c2}(0) \approx 44.4$ T by adding 10 %at DWCNT. This greatly exceeds the upper critical field performance of other intermetallic superconductors such as $Nb_3Sn$, confirming that this compound is very attractive for applications.

**Acknowledgements**

This work was supported in part at Bariloche by CONICET, Fundación Antorchas, and SECYT – PICT, and at Los Alamos by the Office of Energy Efficiency and Renewable Energy, US Department of Energy. The authors are grateful to Judith L. McManus-Driscoll, University of Cambridge, UK, for helpful discussions.

Table I – Sample data for nominal DWCNT %at and actual C content (x). $T_c^{mag}$ and $T_c^{trans}$ were determined from magnetization and transport measurements. The parameters $\eta$ and $g$ were deduced from the fit of $H_{c2}(T)$ curves with equation 1.

| Sample | %at | x | $T_c^{mag}$ | $T_c^{trans}$ | $\eta=D_\pi/D_\sigma$ | g |
|---|---|---|---|---|---|---|
| CNT00 | 0 | 0 | 38.5 | 38.7 | 1 | 0 |
| CNT01 | 1 | 0.015 | 37.5 | 38.4 | 0.1695 | 0.0018 |
| CNT25 | 2.5 | 0.026 | 36 | 37.5 | 0.1288 | 0.0161 |
| CNT05 | 5 | 0.035 | 34 | 35 | 0.1862 | 0.0291 |
| CNT75 | 7.5 | 0.040 | 34 | 35.8 | 0.1409 | 0.0275 |
| CNT10 | 10 | 0.043 | 33.5 | -- | 0.1212 | 0.0355 |
| CNT125 | 12.5 | 0.052 | 31 | 31.1 | 0.2075 | 0.0536 |

**Captions**

Fig. 1. $T_c$ determined from magnetization (solid symbols) and resistivity (open symbols) vs. the actual C content (x) as determined by XRD (see text). The upper inset shows the RRR as a function of x. A typical TEM image of a DWCNT with a 3 nm diameter is displayed in the lower inset (arrows indicate the walls of the DWCNT).

Fig. 2. $J_c$ field dependence determined by magnetization for samples CNT00, CNT25, CNT10 and CNT125, at two temperatures 5K (solid symbols) and 20K (open symbols). The inset illustrates $J_c$ as function of x at 4T and two temperatures, 5K and 20K.

Fig. 3. (a) Transport measurements of the upper critical field ($H_{c2}$) and the beginning of the dissipation ($H_{irr}$) in the R(H) curves for CNT25, CNT10 and CNT125 samples. The inset shows $H_{c2}$ and $H_{irr}$ as a function of x at 4K (squares) and 20K (circles), and the $H_{c2}$

extrapolation at 0K (stars). (b) $H_{c2}$ vs t for CNT00, CNT01 and CNT10 samples and fit to data using equation 1 (dash lines). The inset shows the dependence of $1/D_\pi$ with x (dot line is a guide to the eye).

Figure 1 (A. Serquis et al)

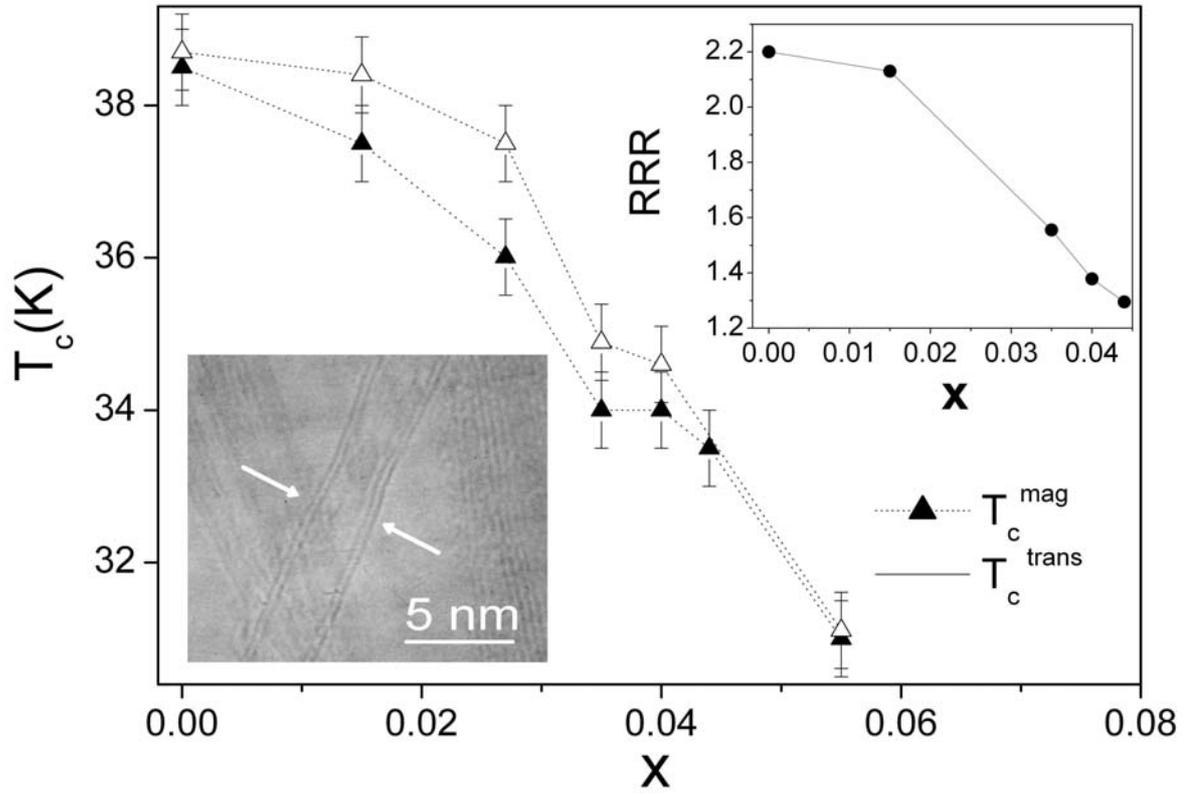

Figure 2 (A. Serquis et al)

Figure 3 (A. Serquis et al)

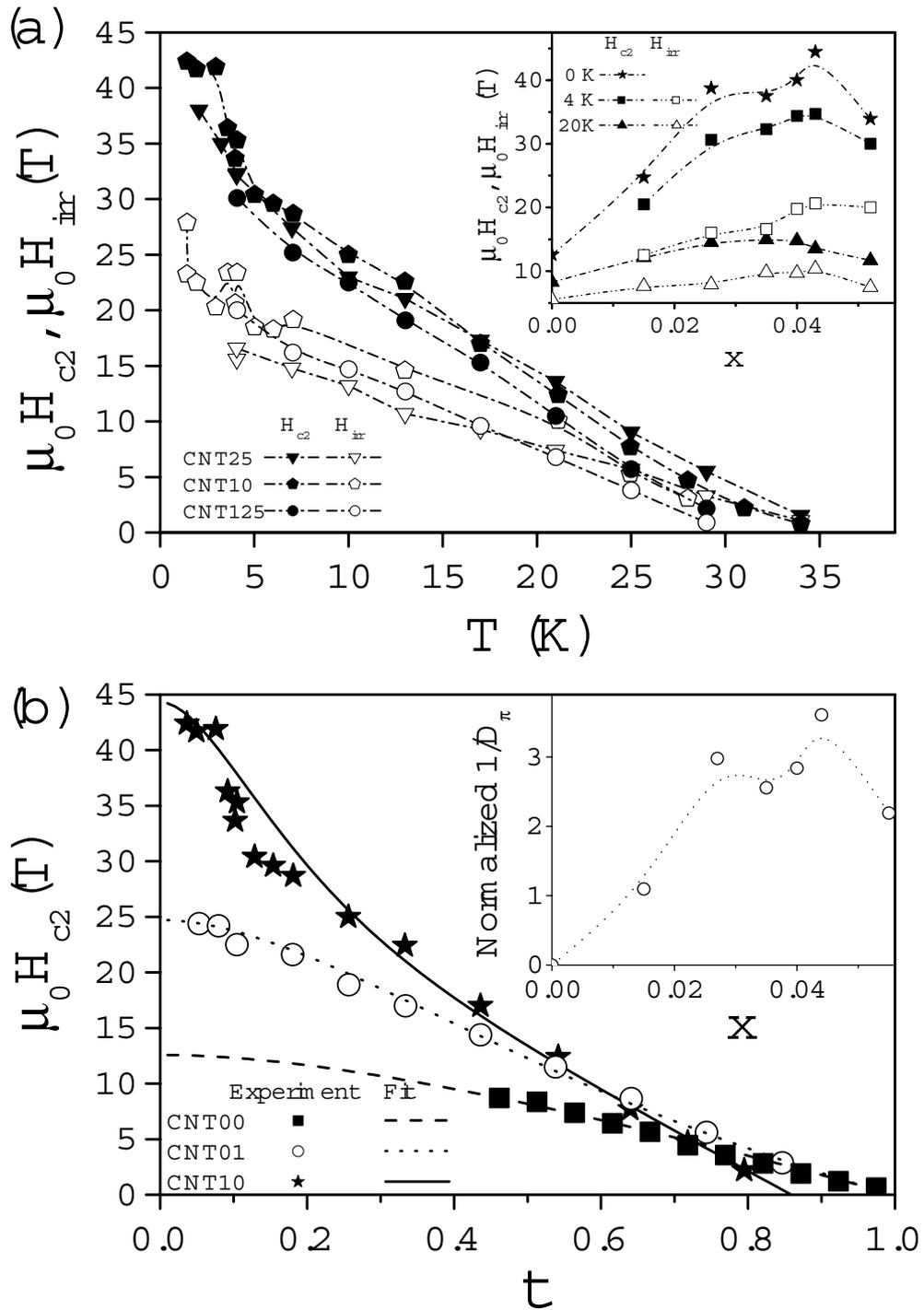